\documentclass[twocolumn, aps, prx, reprint]{revtex4-2}
\usepackage{staples_style}

\begin{document}

\title{Scalable Postselection of Quantum Resources}
\author{J. Wilson Staples\textsuperscript{1}}
\email{will.staples@princeton.edu}
\author{Winston Fu\textsuperscript{1}}
\author{Jeff D. Thompson\textsuperscript{1,2}}
\affiliation{\textsuperscript{1}Princeton Quantum Initiative, Princeton University, Princeton, NJ 08544, USA}
\affiliation{\textsuperscript{2}Department of Electrical and Computer Engineering, Princeton University, Princeton, NJ 08544, USA}

\noindent

\date{\today}

\begin{abstract}

The large overhead imposed by quantum error correction is a critical challenge to the realization of quantum computers, and motivates searching for alternative error correcting codes and fault-tolerant circuit constructions. Postselection is a powerful tool that builds large programs out of probabilistically generated sub-circuits, and has been shown to increase the threshold of quantum error correction based on fusing fixed-size resource states or concatenated codes. In this work, we present an approach to lower the overhead of quantum computing using \emph{scalable postselection}, based on directly postselecting sub-circuits with a size extensive in the code distance using decoder soft information. We introduce a metric, the partial gap, that estimates what the logical gap of a resource state will be after it is consumed, and show that postselection based on the partial gap leads to scalable improvements in the logical error rate. In the specific context of implementing logical gates via teleportation through a cluster state, we demonstrate that scalable postselection provides a $4\times$ reduction in the overhead per logical gate, at the same logical error probability.
    
\end{abstract}

\maketitle

\section{Introduction}

\begin{figure}[t]
    \centering
    \includegraphics[scale=1]{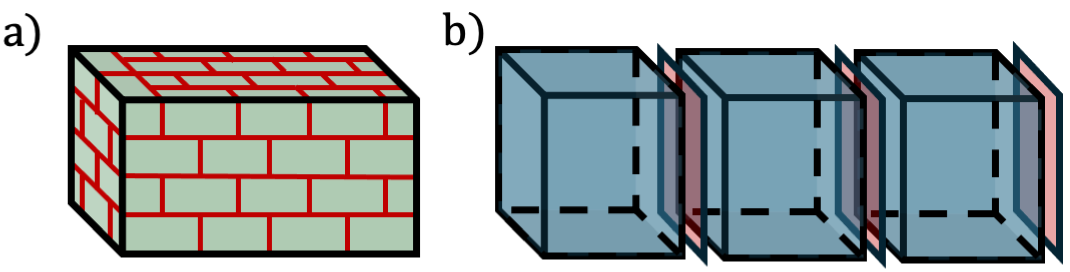}
    \caption{a) Fusion-based computation achieves higher thresholds by constructing circuits by joining fixed-size resource states that are postselected on absence of detected errors. b) Our approach to scalable postselection joins resource states with a size extensive in the code distance that are scalably postselected based on a soft decoder output called the \emph{partial gap}.}
    \label{fig:ps_apps}
\end{figure}

Quantum error correction enables the execution of large quantum programs with realistically noisy hardware~\cite{preskill_reliable_1998}. However, it comes at the expense of significant overhead associated with redundant encoding and fault-tolerant computation. Developing strategies to reduce this overhead in view of realistic hardware constraints is an active area of research~\cite{gottesman_fault-tolerant_2014, xu_constant-overhead_2024, gidney_magic_2024, webster_pinnacle_2026}.

One family of strategies uses postselection to build large programs out of smaller sub-circuits that can be independently retried until they succeed according to some acceptance criterion. By rejecting trials with detected errors, the resulting program has a lower effective error rate. While this can lower the QEC overhead by allowing the use of smaller codes, retries also increase the overhead. Naively, the probability of detecting no errors in a sub-circuit grows exponentially with the size of that circuit. Therefore, achieving practical advantage requires an approach to mitigate the number of retries. One approach is to keep the sub-circuit size constant, as in fusion-based quantum computation (Fig. \ref{fig:ps_apps}a)~\cite{bartolucci_fusion-based_2021, paesani_high-threshold_2023}. As an example, this method can use sub-circuits of only $6$ qubits to achieve a threshold of $11.98\%$ under erasure errors. Another approach is to apply postselection hierarchically in a concatenated code architecture, where the large sub-circuit size at the upper levels of concatenation is offset by an exponentially decreasing error rate~\cite{knillQuantumComputingRealistically2005, aliferis_accuracy_2008}.

An alternative approach is to assemble postselected circuits whose size scales extensively with the QEC code, and instead avoid incurring exponentially many retries by allowing some detected errors and maintaining a fixed rejection fraction. We call this kind of approach {\em scalable postselection}. The feasibility of scalable postselection relies on the notion that logical errors arise almost entirely from atypical events that are readily identifiable~\cite{smith_mitigating_2024-1, chen_scalable_2025, lee_efficient_2025}. Recent work has shown that postselection based on soft decoder output in the form of the logical gap (or complementary gap)~\cite{bombinFaultTolerantPostselectionLowOverhead2024, gidneyYokedSurfaceCodes2025, meisterEfficientSoftoutputDecoders2024} can suppress errors in surface code memory experiments from $\bar{p}$ to $\bar{p}^{ b}$ with $b\approx 2.7$, while maintaining an exponentially small rejection fraction~\cite{chen_scalable_2025}. Similar results have been obtained for general LDPC codes~\cite{lee_efficient_2025}. Demonstrations of postselection based on the logical gap rely on decoding graphs with closed boundaries that allow unambiguous assignment of error strings to logical operators. This is directly useful in state preparation tasks for suppressing errors that are well separated from an open time-like boundary, as found in magic state cultivation~\cite{gidney_magic_2024} or entanglement boosting~\cite{sunamiEntanglementBoostingLowvolume2025a}. When applied to entire circuits, postselection based on the complementary gap can improve the accuracy in applications where extremely low logical failure probability is required~\cite{zhou_error_2025,aharonovSyndromeAwareMitigation2025, dincaErrorMitigationLogical2026}, but it cannot reduce the overhead needed to run a circuit of $N$ gates with a total failure probability $N \bar{p} \approx 1$~\cite{chen_scalable_2025}.

In this work, we propose an approach to reduce the overhead of large algorithms using scalable postselection. The computation proceeds by preparing a series of resource states, measuring their bulk stabilizers for postselection, and teleporting the data through the state after accepting it (Fig. \ref{fig:ps_apps}b). The size of the resource states is extensive in the code distance $d$, so to ensure scalability, we analyze the performance at fixed rejection rate. The rejection decision is based on a quantity called the \emph{partial gap}: the expectation value of the logical gap over the possible values of the unmeasured boundary stabilizers of the resource state. While the exact partial gap is exponentially difficult to compute, we present an approximate method for the surface code that is computationally efficient. For rejection rate $r = 1/2$, we find that the surface code logical error rate is determined by the sum of a bulk and boundary contribution, where the bulk contribution scales as $\bar{p}^{\ b}$ ($b \approx 3/2$) and the boundary contribution includes only errors from the unmeasured boundaries which are subject to a higher threshold. In a context where idle errors are not significant, this approach enables a four-fold reduction in the spacetime overhead required to implement a logical gate with a given error probability.

\begin{figure*}
    \centering
    \includegraphics[scale=1]{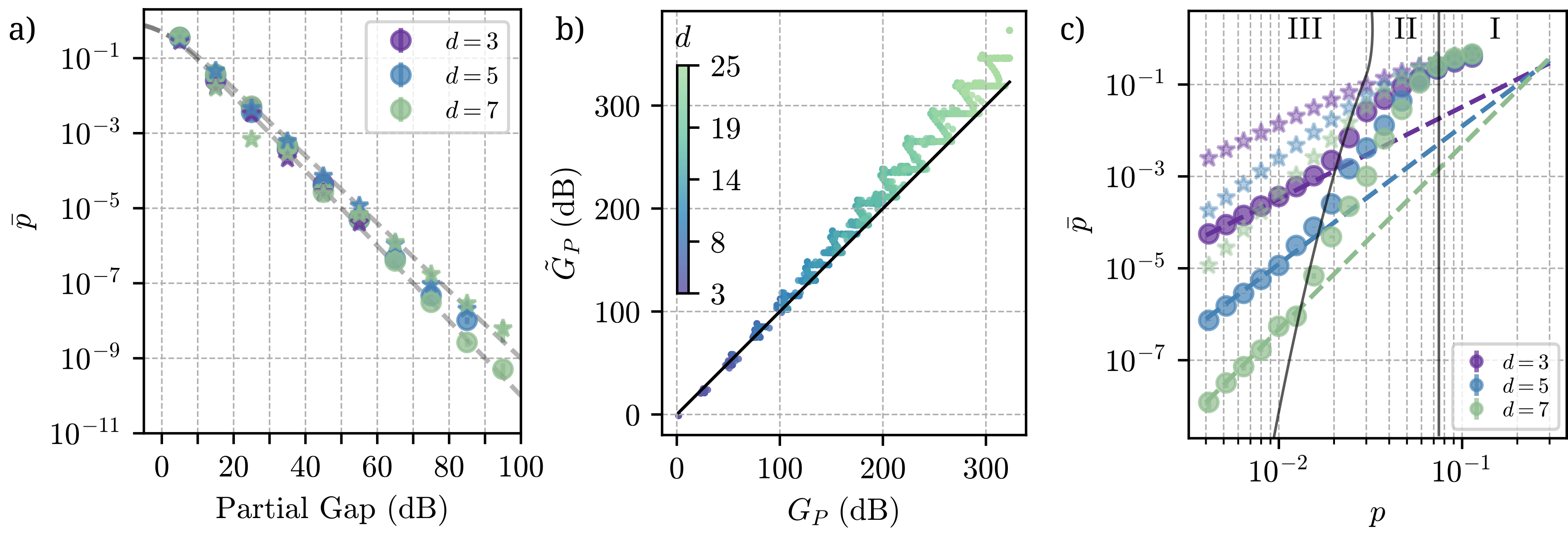}
    \caption{Partial gap for the repetition code. (a) Logical error rate after $d+1$ rounds of syndrome extraction on a distance $d$ repetition code, binned by $G_P$ (circles) or $\tilde{G}_P$ (stars) predicted from the first $d$ layers of syndromes. The dashed lines show logistic models \eqref{eq:logistic} with scaling factors $\alpha \in \{0.9, 1\}$. (b) Scatter plot showing the correlation between $G_P$ and $\tilde{G}_P$ for the repetition code at many distances ($p=10^{-3}$). The solid line shows  $\tilde{G}_P=G_P$. (c) Logical error rate $\bar{p}$ as a function of physical error rate with postselection using $G_P$ at rejection rate $r=1/2$. The error rate with no postselection is shown for reference (stars). Three regimes are indicated schematically denoting above-threshold (I), bulk-error-limited (II) and boundary-error-limited (III). The results are virtually indistinguishable if the approximate partial gap $\tilde{G}_P$ is used instead for postselection (not shown).
    }
    \label{fig:rep-code}
\end{figure*}

\section{Partial Gap}

The logical gap $G$ is defined as the relative probability of the most likely errors corresponding to distinct logical outcomes. For a function $w$, which gives the log-likelihood of the most likely errors conditioned on a measured syndrome $\sigma$ and respective logical outcomes $\ell_0$ and $\ell_1$, we can write the logical gap as~\cite{bombinFaultTolerantPostselectionLowOverhead2024, gidneyYokedSurfaceCodes2025}: 
\begin{equation}
    G(\sigma) := 
    \exp\left(-\left| w(\sigma, \ell_0) - w(\sigma, \ell_1)\right|\right)
\end{equation}
This is a valuable metric for postselection because it is a reliable predictor of the logical error rate, that is, the probability that the decoder picked the wrong logical outcome.

Now consider the case of postselecting a cluster state with $d$ rounds of stabilizer measurements of a distance $d$ surface code~\cite{raussendorf_fault-tolerant_2006}. After preparing the state, the syndromes in the bulk are measured, leaving the initial and final layers unmeasured to teleport data through the state. As a result, the decoding graph used to compute the logical gap has open boundary conditions in the time-like direction. In order to determine a logical gap based on the observed bulk syndrome, it is necessary to make a guess about the eventual value of the boundary stabilizers. However, the value of $G$ is sensitive to this guess, and may not turn out to be a faithful estimator of the logical error rate once the boundary layer is revealed.

To address this, we introduce a different metric, the partial gap $G_P$, which is qualitatively defined as the expectation value of the logical gap taken over different configurations of the unmeasured boundary syndromes.

Explicitly, we create a matching graph including the unmeasured boundary, labeling these checks as the ``hidden syndrome" $\sigma_h$. The bulk syndrome, which has already been revealed after preparing the resource state, is the ``visible syndrome" $\sigma_v$. To compute an expectation value over the hidden syndrome, we need to determine the relative probability of each value of $\sigma_h$, given the already-observed $\sigma_v$. We can estimate this from the weight of the most likely error in each logical class according to:
\begin{equation}
    \PP(\sigma_h | \sigma_v) := \exp\left(-w(\sigma, \ell_0)\right) + \exp\left(-w(\sigma, \ell_1)\right)
\end{equation}
The expectation value of the gap follows as:
\begin{equation}
    \label{eq:pg_def}
    G_P(\sigma_v) := \frac{\sum_{\sigma_h}\PP(\sigma_h | \sigma_v)G(\sigma_v \oplus \sigma_h)}{\sum_{\sigma_h}\PP(\sigma_h | \sigma_v)} 
\end{equation}

Exact computation of $G_P$ is challenging even with an efficient decoder $w$, as the sums contain a number of terms that is exponential in the size of $\sigma_h$. Therefore, we consider several efficiently computable approximations of $G_P$ and study their performance in the context of the repetition code and surface code.

\subsection{Partial gap on the repetition code}

We first study the partial gap for the repetition code, which is simple enough that we can perform a brute-force computation of Eq.~\eqref{eq:pg_def}. We perform a stabilizer simulation for distance $d \in \{ 3, 5, 7\}$ under circuit-level noise with varying strength $p$, using Stim~\cite{gidney_stim_2021}. Each simulation uses a cluster state with $d$ layers of stabilizers followed by a layer of noiseless checks, for a total of $d+1$ layers. All but the noiseless layer are used to calculate $G_P$, then the last layer is re-introduced to determine the logical outcome. The logical error rate is determined by decoding with PyMatching~\cite{higgottSparseBlossomCorrecting2025}, and $G_P$ is computed by direct summation of Eq.~\eqref{eq:pg_def} over all possible boundary configurations, determining $G(\sigma)$ for each using the modified matching approach described in Ref.~\cite{gidneyYokedSurfaceCodes2025}. 

We combine all results from all distances and noise strengths into a histogram showing the dependence of logical error rate $\bar{p}$ on $G_P$ (Fig.~\ref{fig:rep-code}a). We observe agreement with the logistic model:
\begin{equation}
    \label{eq:logistic}
    \bar{p} = \frac{1}{1 + \exp(-\alpha\tilde{G})}
\end{equation}
for some $0.9 \leq \alpha \leq 1$, consistent with literature on the logical gap~\cite{gidneyYokedSurfaceCodes2025}. This indicates that $G_P$ is a reliable estimator of the logical error rate of the full decoding graph after the hidden syndrome is measured.

Motivated by the notion that logical failures primarily result from identifiable rare events, we consider whether the sums in Eq.~\eqref{eq:pg_def} can be approximated by a single term each. Using the same numerical simulation, we compute an approximate gap $\tilde{G}_P$ by implementing a greedy search over $\sigma_h$ to find the most significant contribution to each sum. The result is also shown in Fig.~\ref{fig:rep-code}a, and performs only slightly worse than the brute-force $G_P$. We also compare $G_P$ and its approximation $\tilde{G}_P$ at the level of individual shots (Fig. \ref{fig:rep-code}b), finding generally good agreement across a range of code distances, though $\tilde{G}_P$ often slightly overestimates the confidence in the result.

Finally, we study the logical error rate of the repetition code using $G_P$ as a metric for postselection (Fig. \ref{fig:rep-code}c). Using the same simulation described above, we plot the logical error rate with no postselection, and with postselection based on the partial gap using a fixed rejection rate $r=1/2$. We observe a significant suppression of the logical error rate using postselection based on the partial gap.

Interestingly, the nature of this suppression varies based on the parameter regime, which we label in the plot I-III. Regime I is above threshold and has no error correction. In regime II, the logical failures are dominated by error strings in the bulk. This is similar to full circuit postselection~\cite{smith_mitigating_2024-1, chen_scalable_2025}, and the improvement can be described as an enhancement of the code distance by a factor $b = 1.71 \pm 0.11$. In regime III, logical failures are dominated by errors within the hidden boundary layer. As these errors arise from faults during teleportation, they occur after the postselection decision and cannot be suppressed. However, because this region of the matching graph has lower dimension, errors confined there are governed by a higher effective threshold. By extrapolating data from this regime, we measure an effective threshold $p_{\mathrm{th}} \approx 0.26 \pm 0.013$. This is consistent with the repetition code threshold of 0.5 under code-capacity noise, and the fact that two gates act on each qubit in the circuit-level implementation.

\subsection{Partial gap on the surface code}

The close alignment of $\tilde{G}_P$ with $G_P$ for the repetition code motivates a similar approximation for the surface code. However, a greedy search across the entire space of syndromes on the surface code boundary is also not feasible. In order to find the most important hidden syndromes, we propose a second, heuristic method called ``string splitting". We present the intuition behind this method here, and provide a more exact description in Appendix~\ref{app:string_splitting}. 

\begin{figure*}[t!]
    \includegraphics[scale=1]{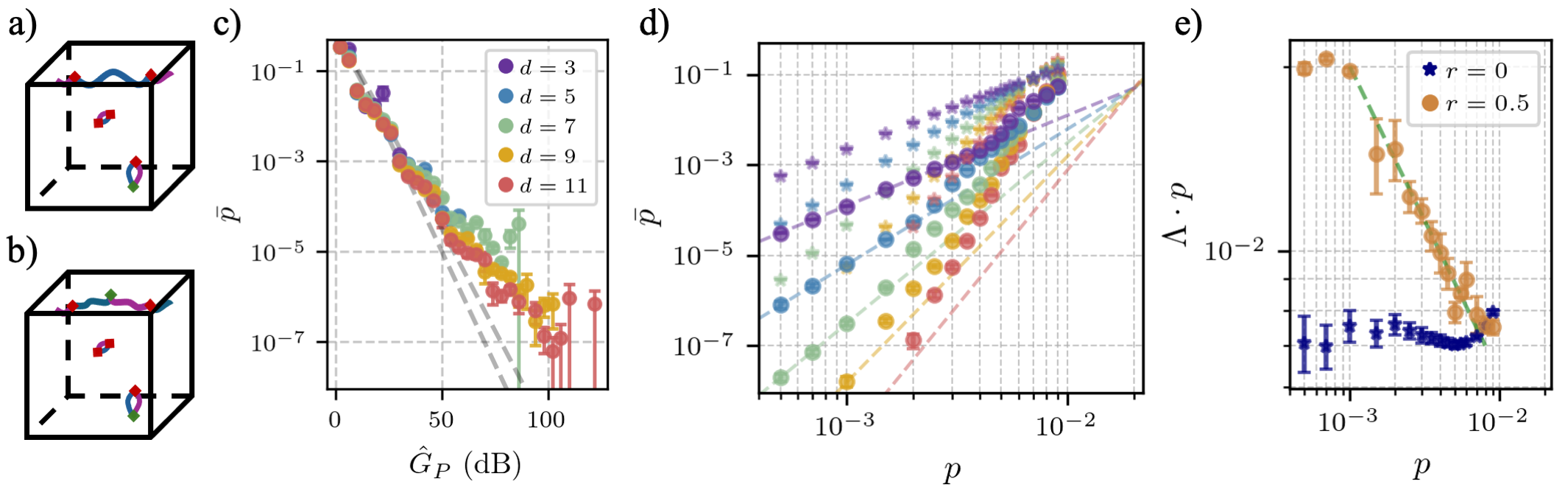}
    \caption{Partial gap for the surface code. (a) A schematic matching graph with visible syndrome $\sigma_v$ (red diamonds), and the most likely hidden syndrome $\sigma_h$ (green diamonds). The most likely corrections for each logical outcome are shown as colored strings (blue, purple). The critical string is the logical operator at the top of the cube which is mostly blue and partially purple. (b) Making an adjustment to $\sigma_h$, shown as a green diamond at the top of the cube, only slightly alters the logical operator. However, its partition into blue and purple sub-strings is greatly affected. (c) Logical error rate against the approximate partial gap $\hat{G}_P$ ($p=2 \times 10^{-3}$). The grey curves show logistic behavior (\eqref{eq:logistic}) for $\alpha = 0.9$ and $\alpha = 1.0$. (d) Logical error rate after $d+2$ rounds of surface code simulation, after postselecting on $\hat{G}_P$ using the inner $d$ layers with $r=1/2$ (circles) and without postselection (stars). The dashed lines show exponential fits to the behavior far below threshold, extrapolating to an effective threshold of $2.01\%$ for boundary errors.  (e) A regression of $\bar{p}$ against $d$ for each value $p$ gives a distance scaling coefficient $\Lambda$. Without postselection, we expect constant $\Lambda \cdot p$. With postselection, we see enhanced $p$-dependence, which we quantify with a power-law fit with slope $m \approx -0.505$ (green line). This implies effective distance augmented by a factor of $1.505$.}
    \label{RSC_post}
\end{figure*}

The string splitting approach is based on searching through a restricted space of hidden syndromes. To accomplish this, we consider the ``critical string" $c$ as the logical operator of minimum gap given a syndrome $\sigma$. We can find $c$ by finding the most-likely error strings conditioned on each logical outcome: $e_0$ and $e_1$. The only non-trivial logical operator supported on the product $e_0e_1$ is $c$. The idea behind string splitting is that, given a $\sigma_h$ and a $c$, we may be able to alter $\sigma_h$ such that $c$ is mostly unchanged but its partition into $e_0$ and $e_1$ changes significantly. If $c$ is unchanged, we know that $w(\sigma, \ell_0) + w(\sigma, \ell_1)$ is also unchanged. Noting:  
\begin{equation}
    \log \left(\PP(\sigma)G(\sigma) \right) \approx  -\max \{w(\sigma, \ell_0), w(\sigma, \ell_1)\} 
\end{equation}
we can see that maximizing $\PP \cdot G$ is accomplished by balancing the contributions of the two logical outcomes. So, we search for adjustments to $\sigma_h$ along $c$ which redistribute errors evenly into each logical class (Fig. \ref{RSC_post}a-b). To accomplish this, we recursively search the region of the boundary corresponding to $c$. We initialize at the value $\sigma_h$ obtained by maximizing $\PP$, then proceed by testing all single bit adjustments. We search deeper for adjustments which improve on $G \cdot \PP$, up to a maximum depth (chosen to be three in this work).

This procedure results in another approximation to the partial gap, $\hat{G}_P$. To validate the performance, we simulate the rotated surface code with distance $d \in \{ 3, 5, 7, 9, 11\}$ under circuit-level noise with varying strength $p$. Each simulation uses a cluster state with $d+1$ layers of stabilizers followed by a layer of noiseless checks. Both the first and last layers of syndromes are hidden during postselection, simulating ambiguity at both time-like boundaries of the resource state. In Fig.~\ref{RSC_post}c, we compare the final logical error rate with approximate partial gap $\hat{G}_P$ for a noise strength $p=0.002$. The approximate partial gap is a faithful predictor of the logical error rate for shots with a high logical error probability, but is less faithful in the low error rate regime. Unlike the repetition code case, we are not able to compute an exact partial gap for comparison.

Next, we study the logical error rate of the surface code using $\hat{G}_P$ as a criterion for postselection (Fig.~\ref{RSC_post}c-d). Using the same simulation described above, we plot the logical error rate with no postselection, and with postselection based on the partial gap using a fixed rejection rate $r=1/2$. We observe a significant suppression of the logical error rate, with the same qualitative regimes as in the repetition code case. Regime II just below threshold is dominated by bulk errors and characterized by a larger effective distance, while regime III is dominated by boundary errors subject to a higher threshold.

Quantitatively, we can extrapolate the effective threshold of the boundary errors in regime III by fitting the dashed curves according to the ansatz $\bar{p} \propto (p/p_{\mathrm{th}})^{\lceil d/2 \rceil}$. We find $p_{\mathrm{th}} = 0.0201 \pm 0.0002$. We physically interpret this as arising from boundary errors within the two-dimensional hidden layer. Accounting for the fact that each qubit in this layer is touched by four noisy gates, this effective threshold is close to the code-capacity surface code threshold with the MWPM decoder of 0.103~\cite{bombin_optimal_2007}.

To quantify the regime II behavior, we compute the scaling of error rate with distance using the metric $\Lambda$: 
\begin{equation}
    \Lambda := \frac{\bar{p}_{d}}{\bar{p}_{d+2}} \approx \exp \left( -2\cdot \partial[\log \bar{p}]/\partial[d] \right)
    \label{Lambda_def}
\end{equation}
Without postselection, we expect $\Lambda \cdot p$ to be approximately independent of $p$, which we observe in Fig.~\ref{RSC_post}e. However, $\Lambda \cdot p$ with postselection varies significantly with $p$, and from a power-law fit with slope $m \approx -0.505$, we infer an effective distance enhancement by a factor $b=1 - m \approx 3/2$. At very low $p$, $\Lambda \cdot p$ is again constant at a value approximately 3 times higher than without postselection, which is another way to observe the higher effective threshold of the boundary errors.

\section{Improvement in Spacetime Overhead}

\begin{figure*}[t!]
    \includegraphics[scale=1]{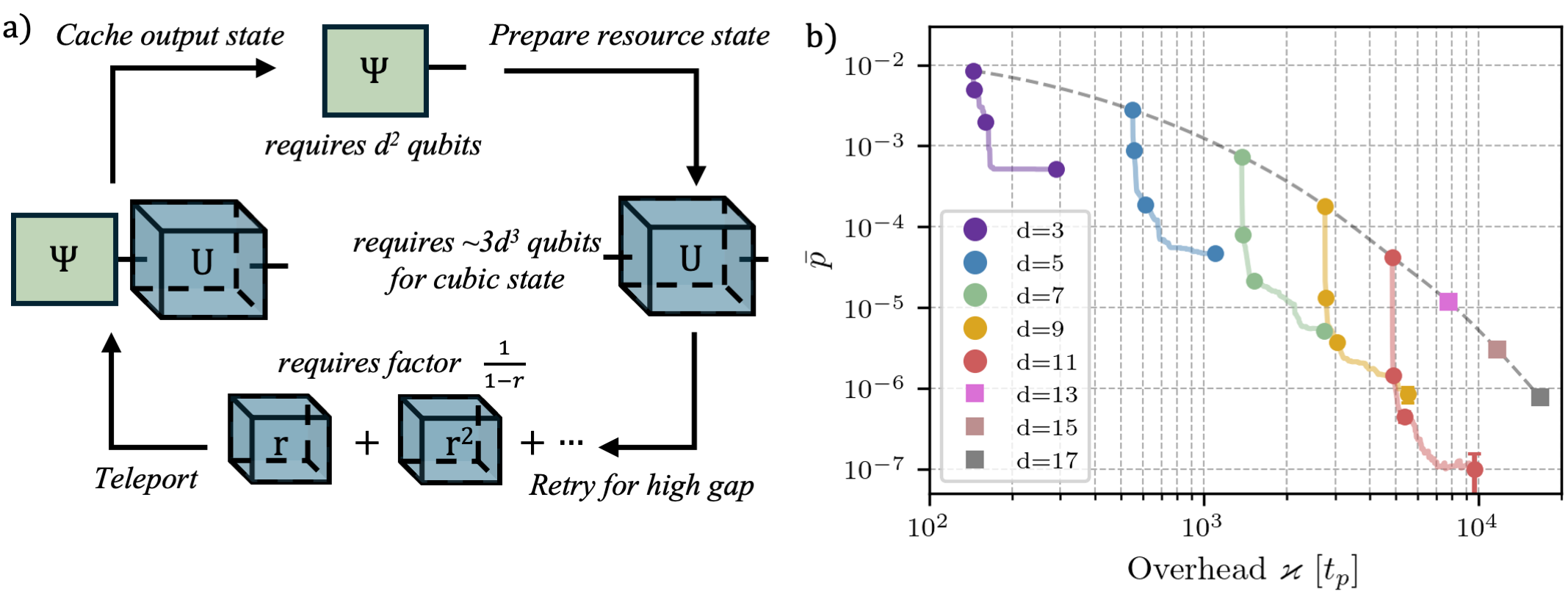}
    \caption{a) An example compute cycle leveraging postselected resource states. The cycle begins with a state $\Psi$ in memory. We then prepare a resource state implementing a single logical gate $\ket{U}$ and $d$ layers of parity checks. The state is rejected with probability $r$, incurring an overhead factor of $(1-r)^{-1}$. After accepting a state, we teleport $\Psi$ through $\ket{U}$ and return it to memory before the next operation. b) Logical error rate per logical gate, as a function of overhead $\varkappa$ for various strategies ($p = 2 \times 10^{-3}$). Colored curves correspond to a given distance, varying the reject rate (dots denote $r \in \{0, 0.01, 0.1, 0.5\}$). Squares show larger distances (without postselection) for reference, along with a spline connecting the $r=0$ points to guide the eye.
    }
    \label{compute_cycle}
\end{figure*}

Distance-parameterized families of error correcting schemes induce a Pareto frontier between resource overhead and logical fault rates. In this paradigm, varying the postselection rate $r$ offers an alternative strategy of trading additional resources for access to lower logical fault rates. In this section, we show that postselection based on the partial gap provides access to an improved frontier of possibilities.

In order to quantitatively explore this, we define the {\em spacetime overhead} $\varkappa$ as a metric resource of consumption. The two expensive aspects of a quantum program are its workspace $q$ of physical qubits and its runtime $t$. Since these can often be traded in a linear manner~\cite{litinski_game_2019-1}, we consider the unified spacetime cost to be $q\cdot t$. For a program with $N$ logical gates, we then define the spacetime overhead of the error-correcting scheme to be $\varkappa := \frac{q\cdot t}{N}$.

To explicitly calculate $\varkappa$, we describe a toy model of gates executed by cyclically preparing and consuming resource states (Fig. \ref{compute_cycle}a). Each cycle begins with a cached logical state $\Psi$ stored in a distance $d$ surface code. We then prepare a resource state with $d$ layers of syndrome measurements, which we equate with executing a single logical gate. We repeat this preparation as necessary to achieve a high partial gap, then teleport $\Psi$ through the resource state. Each preparation step takes time $t_p$. We assume that this timescale is much less than the physical coherence time, so that we may neglect the effect of idling errors on $\Psi$.

This model is periodic, allowing us extract to overhead as an intensive characteristic and measure the trade-off with logical error rate (Fig. \ref{compute_cycle}b). We show different distance protocols as colored curves parameterized by $r\in[0, 1/2]$, using $p=0.002$. For each distance, postselection strictly increases the overhead (by requiring retries), but also decreases the logical error rate. For the range of distances shown, postselection reaches the same logical error rate per gate with approximately 4 times less overhead than a non-postselected strategy. This demonstrates that postselection offers an improved frontier beyond the possibilities with only $d$ as a degree of freedom.

\section{Discussion}

In this paper we present the partial gap and several efficiently computable approximations as confidence metrics for scalable postselection of resource states. We demonstrated that implementing gates by teleportation between postselected circuit chunks enables an approximately four-fold reduction in the overhead necessary to reach a given logical error rate. The practical realization of this approach requires hardware with idle coherence times that are much longer than the time required to prepare, measure and postselect a resource state. Large ratios of coherence time to gate time are often observed in neutral atom, trapped ion and photonic qubits, which may be natural implementation settings.

The quantitative value of the spacetime overhead reduction is sensitive to the strategy that is chosen as a starting point for comparison. In this work, we compare to be implementing gates by teleporting through cluster states using the same approach but without postselection. While the cluster-state approach has a $\sim 1.5-2\times $ higher overhead compared to circuit-based syndrome extraction, it has other advantages in terms of tolerance to qubit loss and structured error propagation that may make it the preferred approach for photonic~\cite{bartolucci_fusion-based_2021} or neutral atom qubits~\cite{sahay_high-threshold_2023} even without postselection.

While we demonstrated that the greedy partial gap $\tilde{G}_P$ is a faithful approximation of $G_P$ for the repetition code, the string splitting approximation $\hat{G}_P$ loses accuracy for shots with small logical error probability. The logical error rate where $\hat{G}_P$ is no longer faithful in Fig.~\ref{RSC_post}c for each distance is approximately the mean logical error rate for that distance, whereas we are relying on the gap to postselect to much lower logical error rates. Therefore, it is possible that a better approximate metric exists. A better metric would not improve the boundary error floor in regime III, but could improve the logical error rate in regime II or increase the acceptance fraction in regime III.

For simplicity, this work considers postselecting circuit chunks with $d$ layers of syndromes, representing the standard resource cost of a single logical gate in the surface code. However, we note that for large code distances, nearly the full benefit of postselection is realized for very small rejection rates, i.e. $r \approx 0.01$ (Fig.~\ref{compute_cycle}b). Therefore, it may be practical to postselect on larger circuit chunks that include multiple logical qubits and logical gates, potentially implemented with transversal gates within the resource state. Extending the partial gap to include $k>1$ logical qubits, and including the effects of correlated decoding~\cite{cain_correlated_2024, serra-peralta_decoding_2026}, is left to future work.

Another natural extension is generalizing the notion of partial gap to other codes. Recent work has made significant headway in creating high-quality confidence metrics for codes beyond the surface code and its matchable relatives~\cite{lee_efficient_2025, xieSimpleEfficientGeneric2026}. Similar to logical gap, these techniques suffer from boundary ambiguity. Analogous methods of considering an expectation over hidden syndromes could allow scalable postselection in these codes, as well.

\textit{Note added}--- While finalizing this manuscript, we became aware of a similar approach to scalable postselection in Ref.~\cite{birchall_macromux_2026}, which proposes a distinct scoring metric for postselecting large states.

\textit{Acknowledgments}--- We acknowledge Hongkun Chen, Frank Zhang, Sarang Gopalakrishnan, David Huse, Jahan Claes and Shruti Puri for helpful discussions. This work was supported by the National Science Foundation (QLCI grant OMA-2120757, and PHY-2047620), the Army Research Office (W911NF-1810215), the Gordon and Betty Moore Foundation (grant DOI 10.37807/gbmf12253), the Office of Naval Research (N00014-20-1-2426), DARPA MeasQuIT (HR00112490363) and the Sloan Foundation. Winston Fu is supported by the A*STAR National Science Scholarship.

\textit{Competing interests}--- J.D.T is a shareholder in Logiqal, Inc.

\bibliography{staples_bib, winston_refs}

\clearpage
\onecolumngrid
\appendix

\section{More on String Splitting}
\label{app:string_splitting}

As discussed in the main body of the text, the fundamental motivation behind the string splitting approximation is that each of the sums in the definition of Partial Gap (Eq. \ref{eq:pg_def}) is dominated by a single value $\sigma_h$. Stated another way, for two sums corresponding to different $\sigma_v$, the sum with a larger maximal term is very likely to be larger. Our task is then to find the value of the hidden syndrome $\sigma_h$ which maximizes each sum. The numerator sum is particularly tricky, since we must simultaneously consider the ambiguity between the logical classes in addition to the total probability of the syndrome configuration.

Our approach to solving this optimization problem is heuristic. It stems from the observation that the dominant case in which 
\begin{equation}
    \arg\max_{\sigma_h} \PP(\sigma_v \oplus \sigma_h) \neq \arg\max_{\sigma_h}\PP(\sigma_v \oplus \sigma_h)\cdot G(\sigma_v \oplus \sigma_h) 
\end{equation}
is when the critical string passes near or through the hidden region of the matching graph. In this case, adjusting the hidden syndrome can repartition the critical string between the logical cosets without affecting the length of the hidden string. As a consequence, the adjustment from $\sigma_h \to \sigma_h'$ has $\PP(\sigma_v\oplus \sigma_h') \approx \PP(\sigma_v \oplus \sigma_h)$. However, the logical gap resulting from the adjustment could be much smaller. 

We leverage this observation into an algorithm by searching through the possible adjustments to the hidden syndrome which are likely to have the desired effect on the critical string. Namely, those hidden syndromes which differ from the detectors in the critical string by only a time coordinate. We call this subset of the hidden syndrome the {\em shadow} of the critical string. For the numerics considered here, the hidden syndrome is restricted to the first and last planes of the detector matching graph, corresponding to time coordinates $t_i$ and $t_f$ respectively. Given a critical string $c$ expressed as a sequence of detector spacetime coordinates, we can define the shadow $S(c)$ as follows: 
\begin{equation}
    \begin{gathered}
        c = \{ (q_1, t_1), \ (q_2, t_2), \dots , (q_n, t_n)\} \\
        S: \ \ c \mapsto \{(q_1, t_i), \ (q_1, t_f), \ (q_2, t_i), \ (q_2, t_f), \dots (q_n, t_i), \ (q_n, t_f) \}
    \end{gathered}
\end{equation}
The critical string $c$ can be found from the sum of matches $m_0$ and $m_1$ corresponding to each logical class. Notably $c$ inherits the dependency of these matches on $\sigma_h$. Strictly speaking, the sum of $m_0$ and $m_1$ (mod $2$) may contain trivial loops as a consequence of equal weight being chosen for the two matches. These should be removed from the critical string. However, deterministic matching algorithms will choose the same string from a set of equal weight choices for both $m_0$ and $m_1$, and no trivial loops will occur. This work uses PyMatching~\cite{higgottSparseBlossomCorrecting2025}, which has this property. 

We can now write the string splitting algorithm as follows: \\
\begin{algorithm}[H]
\caption{String Splitting}
\KwIn{Partial syndrome $\sigma_{v}$, max depth $D$}
\KwOut{Partial Gap Estimate $\tilde{G}_P$}

\SetKwFunction{ShadowDAG}{SplitString}
\SetKwFunction{EstimateContribution}{EstimateContribution}
\SetKwFunction{AssumeLogical}{GetOpenBoundary}
\SetKwFunction{InferFull}{InferSyndrome}
\SetKwFunction{FindCritical}{FindCriticalString}
\SetKwFunction{GetML}{GetMostLikelySyndrome}
\SetKwFunction{Match}{Match}
\SetKwProg{Fn}{Function}{:}{}

\BlankLine
$\PP, \hat{\sigma} \gets \GetML(\sigma_v) $\;
$\hat{G}_P \gets$ \ShadowDAG{$\hat{\sigma}, D$} $- \log\PP$\;

\BlankLine
\Fn{\ShadowDAG{$\hat{\sigma}$, $D$, $t_0 = \infty$, $d = 0$}}{
    $m_0 \gets \Match(\hat{\sigma}, \ell=0)$, $m_1 \gets \Match(\hat{\sigma}, \ell=1)$\;
    $t \gets \max(w(m_0), w(m_1))$\;
    \eIf{$t > t_0$ OR $d =D$}{
        \Return $\max(t, t_0)$
    }{
    $\mathcal{C} \gets \FindCritical(m_0, m_1)$\;
    $T \leftarrow [ \ ]$\;
    \ForEach{$j \in S(\mathcal{C})$}{
        $\hat{\sigma}' \leftarrow \hat{\sigma} \oplus \mathbf{e}_j$\;
                   Append \ShadowDAG{$\hat{\sigma}', D, t, d+1$} to $T$\;   
        }
    \Return $\min\bigl(t_0,\; \min(T)\bigr)$\;
    }    
}
\end{algorithm}

All numerics presented here use maximal depth $D=3$, since this heuristically saturates performance while maintaining reasonable runtime. 

\section{Simulation Details}
\label{app:sim_deets}

The simulations presented in this work use stim~\cite{gidney_stim_2021} to implement Knill syndrome extraction on the Rotated Surface Code (RSC) to generate our cluster states. Instances of a distance $d$ RSC consist of $d^2$ data qubits and use $d+2$ rounds of detectors ($d$ visible and $2$ hidden) in all cases. The full circuit was created by foliating the following circuit block for each round of desired detectors: 
\begin{center}
    \begin{quantikz}
        \lstick{  } & \qw & \qw & \qw & \ctrl{1} & \gate{\M_X^{\otimes n}} \\
        \lstick{$\ket{+}^{\otimes n}$} & \gate{\M_{\S_Z}} & \ctrl{1} & \qw & \targ{} & \gate{\M_Z^{\otimes n}} \\
        \lstick{$\ket{0 \ }^{\otimes n}$} & \gate{\M_{\S_X}}   & \targ{} & \qw & \qw & \qw \\
    \end{quantikz}
\end{center}
where $\M_\S$ represents syndrome measurement using ancillae and entangling gates. The standard ``NZ" entangling gate pattern was used~\cite{tomita_low-distance_2014}. 

The only entangling gate used in the circuit realization was the control-$Z$. This is relevant because it allows the simplified noise model of dephasing with strength $p$ concurrent with the action of all control-$Z$ gates. The only exception to this rule is the initial ``gauge measurement" of the stabilizer generators and the final round of syndrome extraction, which are both exempt from dephasing. 

The choice to simulate a single resource state preparation rather than several in sequence is consequential since it risks underestimating logical error for failing to consider logical errors that emerge from the interaction of adjacent blocks. Said another way, logical operators which pass {\em through} hidden layers of the detector matching graph, involving edges that join both adjacent layers, are not considered. The reasoning behind this choice is that it is unclear how partial gaps are distributed given a syndrome history. Thus, we cannot easily define a policy about which to accept and which to reject. 

To illustrate this point, consider a simulation of two consecutive teleportations, which we postselect independently. For the first of these, we calculate an approximate partial gap $\tilde{G}_1(\sigma_1)$. Since we make this calculation many times, we have a reliable picture cumulative distribution for $\tilde{G}_1$, and thus we can effectively postselect at a target retry rate $r$. For the second chunk, we have committed to the first syndrome $\sigma_1$, so we calculate the approximate partial gap $\tilde{G}_2(\sigma_1, \sigma_2)$, but we are unlikely to have ever seen $\sigma_1$ before so the cumulative distribution of $\tilde{G}_2$ conditioned on $\sigma_1$ is unknown. Thus, we cannot define a policy for whether to retry, since we are unsure of the probability of improvement with a new $\sigma_2$. 

\end{document}